\documentclass[aps,prb,amsmath,twocolumn,superscriptaddress,showpacs]{revtex4}

\usepackage{graphicx}
\usepackage{bm} 
\usepackage{url}

\DeclareMathOperator{\Img}{\mathrm{Im}}

\DeclareMathOperator{\Sp}{\mathrm{Sp}}
\begin{document}
\author{Mikhail S. Kalenkov}
\affiliation{I.E. Tamm Department of Theoretical Physics, P.N. Lebedev Physical Institute, 119991 Moscow, Russia}
\author{Andrei D. Zaikin}
\affiliation{Institut f\"ur Nanotechnologie, Karlsruher Institut f\"ur Technologie (KIT), 76021 Karlsruhe, Germany}
\affiliation{I.E. Tamm Department of Theoretical Physics, P.N. Lebedev Physical Institute, 119991 Moscow, Russia}

\title{Diffusive superconductors beyond Usadel approximation:
Electron-hole asymmetry and large photoelectric effect}

\begin{abstract}
We extend the quasiclassical formalism for diffusive superconductors by deriving anisotropic (gradient) corrections to the Usadel equation. We demonstrate that in a number of physical situations such corrections may play a crucial role being responsible for the effects which cannot be recovered within the standard Usadel approximation. One of them is the so-called photoelectric effect in superconductors and superconducting-normal (SN) hybrid structures. Provided a superconducting part of the system is irradiated by an external ac electromagnetic field the charge imbalance develops and a non-vanishing dc voltage is induced across the SN interface. In the presence of magnetic impurities in a superconductor the magnitude of this effect becomes large and can easily be detected in modern experiments.
\end{abstract}

\pacs{74.20.-z, 74.25.F-, 74.25.fg}
\maketitle

\section{Introduction}
\label{sec-intro}

Quasiclassical theory of superconductivity \cite{Eilen68,Larkin68,Usadel70,bel} is a powerful tool for comprehensive analysis of the superconducting state. Quasiclassical methods are well suited to study a broad range of phenomena with characteristic energies/frequencies well below
the Fermi energy $\varepsilon_F$ and characteristic lengths strongly exceeding the Fermi wavelength or, equivalently, the inverse Fermi momentum $1/p_F$. Within this approach all relevant information about the superconducting
state can be extracted from the energy-integrated matrix Green-Eilenberger function $\check g$ which generally depends on the electron momentum
direction as well as on both coordinate and time variables.

The quasiclassical Eilenberger formalism also accounts for electron scattering at various impurities and boundary imperfections which are inevitably
present in virtually any metallic sample. In general such scattering processes can make the quasiclassical electron trajectory in a metal very
complicated in which case the function $\check g$ may become rather involved. Usadel \cite{Usadel70} demonstrated that substantial simplification can
be achieved provided the concentration of non-magnetic impurities
in a system is sufficiently large. In this case electron scattering at such impurities turns its trajectory into a diffusive one and the matrix function $\check g$ becomes almost isotropic in the momentum space.  The Eilenberger equation then
reduces to the Usadel one for the momentum averaged quasiclassical Green function matrix \cite{Usadel70,bel}.

Usually the Usadel approximation works well in the so-called dirty limit, i.e. provided the electron elastic mean free path $\ell$
remains much shorter than the superconducting coherence length of a clean system $\xi_0$. However, there exists a number of interesting phenomena
which cannot be captured within the Usadel approximation even in the limit $\ell \ll \xi_0$. One of them is the thermoelectric effect in superconductors
doped with magnetic impurities \cite{Kalenkov12} as well as in superconducting hybrids with spin-active interfaces \cite{KZ14,MISM14,KZ15}.
Different scattering rates for electrons and holes at magnetic impurities or interfaces yield electron-hole imbalance generation in such
systems which in turn results in a large enhancement of thermoelectric currents. This electron-hole imbalance effectively implies anisotropy of the
quasiparticle distribution function in the momentum space \cite{Kalenkov12,KZ14,MISM14,KZ15}. It turns out that this effect cannot be recovered from the standard
Usadel equation for the momentum-averaged quasiclassical Green functions. One of the aims of this paper is to extend further the Usadel perturbative
expansion in the parameter $\ell/\xi_0$ and to derive the leading order correction to the Usadel equation which would enable one to properly account
for the momentum anisotropy of the distribution function mentioned above.

Another our aim is to employ this formalism in order to study the so-called
photoelectric effect in superconducting-normal (SN) hybrid structures. The essence
of this effect is the appearance of a dc voltage drop across a superconductor as a result
of its inhomogeneous absorption of light. It is
commonly accepted that an external electromagnetic ac field cannot cause any charge
imbalance in a superconductor \cite{Eliashberg71}: Photon absorption just produces
a pair of electron-like and hole-like excitations and does not
lead to any shift of the chemical potential. This observation remains applicable
as long as both the density of states and quasiparticle
scattering rates are symmetric with respect to electrons and holes. Electron-hole imbalance occurs provided one takes into account
the energy dependence of the normal state density of states near the Fermi
energy \cite{Gulian96}. The corresponding
effect is relatively small for generic metals without anomalies in the density of
states in the vicinity of the Fermi level. Accordingly, the magnitude of the
photoelectric effect in conventional superconductors is expected to be small in the parameter $T_c/\varepsilon_F$, where $T_c$ is a critical temperature of a superconductor.

The situation changes dramatically if one considers superconductors doped with randomly distributed magnetic impurities.
In this case scattering of quasiparticles on such impurities may lead to significant electron-hole asymmetry resulting, in turn,
in a strong enhancement of the photoelectric effect \cite{Zaitsev82,Zaitsev86} which amplitude is not anymore controlled by the small
parameter $T_c/\varepsilon_F$. Interestingly enough, it turns out that this large photoelectric effect cannot be captured within the usual Usadel
equation formalism. In order to correctly describe this effect in diffusive superconductors doped with magnetic impurities it is necessary to go
beyond the standard Usadel approximation and to include anisotropic (gradient) corrections. This will be demonstrated below in our paper.

The structure of the paper is as follows. In section \ref{sec-quas} we will extend the Usadel expansion of the
quasiclassical Green-Eilenberger functions and derive the leading order correction to the Usadel equation describing superconductors
in the diffusive limit. In section \ref{sec-thermo} we will demonstrate the advantages of our approach by briefly
re-deriving the expression for large thermoelectric currents which may flow in superconductors doped with magnetic impurities.
In section \ref{sec-photo} we will apply our formalism in order to analyze the photoelectric effect in SN structures with magnetic impurities.
Our main conclusions are briefly presented in section V. Some technical details of our calculation are displayed in Appendix.

\section{Quasiclassical description of diffusive superconductors}
\label{sec-quas}

\subsection{Eilenberger formalism}
\label{subsec-eilen}

Our starting point is the quasiclassical Eilenberger equation which can be expressed in the form
\begin{equation}
i\bm{v}_F \bm{\partial}_{\bm{r}} \check g +
\left[\check \Omega - \check \Sigma_{\text{nm}}, \check g\right] =0,
\label{eilen}
\end{equation}
where $\bm{v}_F = \bm{p}_F/m$ is the Fermi velocity, $\bm{\partial}_{\bm{r}}$ denotes spatial gradient and $\check g(\varepsilon, t, \bm{p}_F, \bm{r})$ is the quasiclassical matrix propagator which has the standard structure in the Keldysh space
 \begin{equation}
\check g =
\begin{pmatrix}
\hat g^R & \hat g^K \\
0 & \hat g^A
\end{pmatrix}
\label{green}
\end{equation}
and obeys the normalization condition
\begin{equation}
\check g^2=1.
\label{norm}
\end{equation}
Here and below the ``check''-symbol
denotes $8\times8$ matrices in Keldysh$\otimes$Nambu$\otimes$Spin space while the``hat''-symbol labels $4\times4$ matrices in Nambu$\otimes$Spin space. The brackets in Eq. \eqref{eilen} and below imply the commutator $\left[a, b\right]=ab-ba$.
The matrix $\check \Omega$ has the form
\begin{equation}
\check \Omega = \varepsilon \hat \tau_3 + \check \Delta - \check \Sigma_{\text{m}},
\label{Omega}
\end{equation}
where $\varepsilon$ is the quasiparticle energy, $\hat \tau_3$ is the Pauli matrix in the Nambu space and $\check \Delta$
is the order parameter matrix which contains only ``retarded'' and ``advanced'' components,
\begin{equation}
\check \Delta =
\begin{pmatrix}
\hat \Delta & 0\\
0 & \hat \Delta
\end{pmatrix}, \quad \hat \Delta
=
\begin{pmatrix}
0 & \Delta \sigma_0 \\
-\Delta^* \sigma_0 & 0
\end{pmatrix},
\label{chde}
\end{equation}
where $\sigma_0$ is the unity matrix in the spin space and $\Delta$ is the BCS order parameter in our system. The two self-energies
$\check \Sigma_{\text{m}}$ and $\check \Sigma_{\text{nm}}$ account for electron scattering on respectively magnetic and non-magnetic
impurities contained in our  superconductor. In the case of isotropic elastic impurity scattering the self-energy $\check \Sigma_{\text{nm}}$ reads
\begin{equation}
\check \Sigma_{\text{nm}}
= - i \dfrac{v_F}{2\ell}\left<\check g\right>,
\end{equation}
where angular brackets $\left< \cdots\right>$ denote averaging over the Fermi surface. The expression for $\check \Sigma_{\text{m}}$ will be specified
below. For the sake of simplicity we will assume that the matrix $\check \Omega$ is independent on the momentum direction $\bm{p}_F$.

Electric current density $\bm{j}$ in our system is determined via the Keldysh part of the quasiclassical Green function
\begin{equation}
\bm{j}=
-\dfrac{e N_0}{8}
\int d \varepsilon
\left<
\bm{v}_F
\Sp\left[\hat g^K \hat \tau_3 \right]
\right>,
\label{current-eil}
\end{equation}
where $e$ is the electron charge and $N_0$ is the normal density of states at the Fermi level.

\subsection{Expansion in spherical harmonics}
\label{subsec-harmonics}

Let us now assume that the electron elastic mean free path in a superconductor is shorter than its ``clean limit'' coherence length $\ell \ll \xi_0\sim v_F/\max(|\Delta|, T)$. In this case the electron motion becomes diffusive and the Green-Eilenberger function $\check g$ only weakly depends on the momentum direction $\bm{p}_F$. In order to account for this dependence one can formally expand the function $\check g$ in a series of spherical harmonics
\begin{equation}
\check g = \check g_0  + \check g_i n_i +
\check g_{ij} \left( n_i n_j- \dfrac{1}{3}\delta_{ij}\right) + \cdots,
\label{expansion}
\end{equation}
where $\bm{n} = \bm{p}_F/p_F$ is the unity vector in the direction of the
quasiparticle momentum and the matrices $\check g_0$, $\check g_i$, $\check g_{ij}$ do not depend on the electron momentum.
Here and below the summation over  repeated indices is implied.
As it will become clear below, an effective expansion parameter in Eq. \eqref{expansion} is $\sqrt{\ell/\xi_0}$, i.e.
\begin{equation}
\check g_0 \sim \left(\dfrac{\ell}{\xi_0}\right)^0,
\quad
\check g_i \sim \left(\dfrac{\ell}{\xi_0}\right)^{1/2},
\quad
\check g_{ij} \sim \left(\dfrac{\ell}{\xi_0}\right)^1.
\label{lxi0}
\end{equation}

The standard Usadel approximation \cite{Usadel70} amounts to keeping only the first two terms in the expansion \eqref{expansion}.
Under this approximation the Eilenberger equation \eqref{eilen}  reduces to the Usadel one \cite{Usadel70,bel}. Here we go beyond this approximation and also keep the next order term in Eq. \eqref{expansion} which contains the matrix $\check g_{ij}$.
Without any loss of generality, we can assume that this matrix $\check g_{ij}$ is symmetric and traceless, i.e.
\begin{equation}
\check g_{ij} = \check g_{ji}, \quad \check g_{ii}=0.
\end{equation}

In order to proceed it is convenient to rewrite gradient term in the Eilenberger equation \eqref{eilen} in the operator form as $\bm{\partial}_{\bm{r}} \check g = \nabla \check g - \check g \nabla$. Then Eq. \eqref{eilen} reads
\begin{equation}
\left[\check \Omega + i\bm{v_F}\nabla - \check \Sigma_{\text{nm}}, \check g\right] =0.
\label{eilen2}
\end{equation}
Substituting the expansion \eqref{expansion} into Eq. \eqref{eilen2} and collecting all terms associated respectively with harmonics $1$, $n_i$ and $ n_i n_j- \delta_{ij}/3$, we arrive at the following system of equations
\begin{gather}
\left[
\check \Omega, \check g_0\right]
+
\dfrac{iv_F}{3}
\left[\nabla_i, \check g_i\right]=0,
\label{eilen_0}
\\
\left[
\check \Omega + i \dfrac{v_F}{2\ell} \check g_0,
\check g_i\right]
+
i v_F
\left[\nabla_i, \check g_0\right]
+
\dfrac{2 i v_F}{5}
\left[\nabla_k, \check g_{ik}\right]
=0,
\label{eilen_i}
\\
\left[ \check g_0,
\check g_{ik}\right]
+\ell
\left(
\left[\nabla_i, \check g_k\right]
+
\left[\nabla_k, \check g_i\right]
-
\dfrac{2}{3}\delta_{ik}
\left[\nabla_n, \check g_n\right]
\right)
=0.
\label{eilen_ij}
\end{gather}
Analogously from the normalization condition \eqref{norm} we obtain
\begin{gather}
\check g_0^2 + \dfrac{1}{3}\check g_i \check g_i =1,
\label{norm_0}
\\
\check g_0 \check g_i + \check g_i \check g_0 +
\dfrac{2}{5}(\check g_{ik} \check g_k+ \check g_k \check g_{ik}) =0,
\label{norm_i}
\\
\check g_0 \check g_{ik} + \check g_{ik} \check g_0  +
\dfrac{1}{2}\left( \check g_i  \check g_k + \check g_k \check g_i
- \dfrac{2}{3} \delta_{ik} \check g_n  \check g_n \right)=0.
\label{norm_ij}
\end{gather}
For the sake of brevity here we neglected the vector potential which is formally set equal to zero $\bm{A}=0$. If needed, the vector potential can easily be restored by gauge invariance arguments
implying the replacement
\begin{equation}
\nabla \rightarrow \nabla +i \dfrac{e}{c}\bm{A}(\bm{r},t)\hat \tau_3,
\end{equation}
where $c$ is the speed of light.

In order to evaluate $\check g_{ij}$ it is sufficient to find $\check g_i$ only in the lowest order in the parameter $\ell/\xi_0$. From Eqs. \eqref{eilen_i} and \eqref{norm_i} we get
\begin{equation}
\check g_i = -\ell
(\check g_0 \nabla_i \check g_0 - \nabla_i)
\label{g_iUsadel}
\end{equation}
which is just the well known expression for the amplitude of the first harmonics within the standard Usadel approximation \cite{Usadel70}. Combining this expression with Eqs. \eqref{eilen_ij} and \eqref{norm_ij}, we obtain
\begin{multline}
\check g_{ik} =-
\dfrac{\ell^2}{4}
\biggl[
\nabla_i \nabla_k \check g_0
+
\nabla_k \nabla_i \check g_0
+
\check g_0 \nabla_i \nabla_k
+
\check g_0 \nabla_k \nabla_i
+\\+
\nabla_i \check g_0 \nabla_k
+
\nabla_k \check g_0 \nabla_i
-
3 \check g_0 \nabla_i \check g_0 \nabla_k \check g_0
-
3 \check g_0 \nabla_k \check g_0 \nabla_i \check g_0
+\\+
\dfrac{2}{3}\delta_{ik}
\left(
3 \check g_0 \nabla_n \check g_0 \nabla_n \check g_0
-
\nabla_n \check g_0 \nabla_n
-
\nabla_n \nabla_n \check g_0
-
\check g_0 \nabla_n \nabla_n
\right)
\biggr].
\label{gik}
\end{multline}
For the sake of generality here and below we make a distinction between the products $\nabla_i \nabla_k$ and $\nabla_k
\nabla_i$ being different from each other in the presence of the magnetic field.

\subsection{Correction to Usadel equation}
\label{subsec-corr}

From Eqs. \eqref{norm_0} and \eqref{g_iUsadel} we observe that the normalization condition for the isotropic part of the function $\check g_0$ acquires a nontrivial correction
\begin{multline}
\check g_0^2 = 1- \dfrac{1}{3}\check g_i \check g_i
=
1
-
\dfrac{\ell^2}{3}
\bigl(
\check g_0 \nabla_i \nabla_i \check g_0 + \nabla_i \nabla_i
-\\-
\check g_0 \nabla_i \check g_0 \nabla_i
-
\nabla_i \check g_0 \nabla_i \check g_0
\bigr).
\end{multline}
At this stage it is convenient for us to introduce the function $\check G$ related to $\check g_0$ as
\begin{equation}
\check g_0 = \check G
-\dfrac{\ell^2}{6}
\left(
\check G \nabla_i \nabla_i + \nabla_i \nabla_i \check G
-
2 \nabla_i \check G \nabla_i
\right).
\label{G11}
\end{equation}
It is easy to verify that this function obeys the standard normalization condition
\begin{equation}
\check G^2=1.
\label{Gnorm}
\end{equation}
Combining Eqs. \eqref{eilen_i}, \eqref{norm_i}, \eqref{gik} and \eqref{G11} we also express the matrix $\check g_i$ via $\check G$ and find
the correction to this matrix. The corresponding rather lengthy expression is displayed in Appendix, see Eq. \eqref{giCorr}.
Substituting this expression into Eq. \eqref{eilen_0} we arrive
at the result \eqref{UsadelCorr} which extends the Usadel equation \cite{Usadel70,bel} by including terms describing weak anisotropy effects in the momentum space.

Though Eq. \eqref{UsadelCorr} still looks complicated, in the majority of cases it can be simplified further. For instance, in the absence of the magnetic field
($\bm{A} = 0$) and provided the matrix $\check G$ depends only on one coordinate ($x$) Eq. \eqref{G11} reduces to
\begin{equation}
\check g_0 = \check G
-\dfrac{\ell^2}{6}
\partial^2_x \check G,
\label{g0norm-1}
\end{equation}
while Eq. \eqref{UsadelCorr} becomes
\begin{gather}
\begin{split}
iD\partial_x (\check G \partial_x \check G)
=&
\left[ \check \Omega, \check G \right]
\\-&
\dfrac{\ell^2}{6}
\left[
\partial^2_x \check \Omega, \check G \right]
-
iD\dfrac{\ell^2}{30}
\left[ \partial^4_x \check G
, \check G \right],
\end{split}
\label{usadel-1}
\end{gather}
where $D=v_F\ell/3$ is diffusion coefficient. The first line of Eq. \eqref{usadel-1} represents the standard Usadel equation while its second line defines the leading order gradient corrections to this equation.

The $x$-component of the current density reads (cf. Eq. \eqref{current-us})
\begin{equation}
j_x=
-\dfrac{\sigma_N}{16 e \ell}
\int d \varepsilon
\Sp\left(\hat g_x^K \hat \tau_3 \right),
\label{current-us1}
\end{equation}
where $\sigma_N = 2e^2N_0D$ is the normal state Drude conductivity and $\hat g_x^K$ is the Keldysh component of the matrix \begin{equation}
\begin{split}
\check g_x
=
-&\ell \check G \partial_x \check G
-
i\dfrac{\ell^2}{v_F}
\left[
\check \Omega,  \partial_x \check G
\right]
\\-&
D\dfrac{\ell^2}{20v_F}
\bigl(
3 [\check G,  \partial^3_x \check G]
+
7 [\partial_x \check G,  \partial^2_x \check G]
\bigr).
\end{split}
\label{gx-1}
\end{equation}
The expression \eqref{gx-1} follows directly from Eq. \eqref{giCorr}.

It is important to emphasize that the above results remain applicable as long as significant changes of the function $\check G$ occur at distances exceeding the electron elastic mean free path $\ell$. Should this condition be violated (as it is the case, e.g., in diffusive hybrid structures with spin-active interfaces \cite{KZ15}) the whole approach based on the expansion \eqref{expansion} may become insufficient. On the other hand, usually the function $\check G$ in diffusive superconductors changes at distances of order of the "dirty limit" coherence length
$\xi \sim \sqrt{\xi_0\ell}$ (which also follows from the Usadel equation) or even at longer distances. Hence, the effective expansion parameter in Eq. \eqref{expansion} is
$\ell /\xi \sim (\ell /\xi_0)^{1/2}$ as we already indicated above in Eq. \eqref{lxi0}. Further simplifications occur provided $\check G$ changes at very long distances, in which case the terms with
higher order gradients in Eqs. \eqref{usadel-1} and \eqref{gx-1} can be neglected.

\section{Thermoelectric current}
\label{sec-thermo}

Let us now turn to practical calculations employing the above formalism. Our first example is the thermoelectric effect in superconductors. If a non-vanishing temperature gradient $\nabla T$ is applied to a superconducting sample quasiparticle distribution function is driven out of equilibrium and the quasiparticle current develops in the system. As a result, the total current density consists of two contributions:
\begin{equation}
\bm{j}(\bm{r}) = \bm{j}_s(\bm{r}) + \alpha(\bm{r}) \bm{\partial}_{\bm{r}} T(\bm{r}),
\end{equation}
where $\bm{j}_s(\bm{r})$ defines the supercurrent density which we are not interested in here and the last term represents the thermoelectric current.
For simplicity below we restrict ourselves to the linear response regime, i.e. we assume that both the supercurrent and the thermoelectric current remain small as compared to the critical (depairing) current. In this case the thermocurrent is proportional to $\nabla T$ and $\alpha$ is the so-called thermoelectric coefficient.

Without loss of generality we can orient the $x$-axis in the direction of non-zero temperature gradient, i.e. we assume $T=T(x)$. Employing Eqs. \eqref{usadel-1}, \eqref{current-us1} and \eqref{gx-1} we directly evaluate the thermoelectric current. Provided the temperature varies at a typical length scale strongly exceeding the superconducting coherence length $\xi$ the terms with spatial gradients in  Eq. \eqref{usadel-1} can be safely neglected as they could only yield small corrections to the Green function proportional to $(\partial_x T)^2$ or $\partial_x^2 T$. As a result, it suffices to solve Eqs. \eqref{usadel-1} within the approximation assuming local equilibrium in our superconductor. Then both the Green-Keldysh function and the self-energy take a simple form
\begin{gather}
\hat G^K(\varepsilon, x) =
\left[\hat G^R(\varepsilon, x) - \hat G^A (\varepsilon, x)\right]
\tanh\dfrac{\varepsilon}{2T(x)},
\label{gk-thermo}
\\
\hat \Omega^K(\varepsilon, x) =
\left[\hat \Omega^R(\varepsilon, x) - \hat \Omega^A (\varepsilon, x)\right]
\tanh\dfrac{\varepsilon}{2T(x)},
\end{gather}
where the retarded and advanced Green functions obey Eq. \eqref{usadel-1} which now reduces to
\begin{equation}
\left[\hat \Omega^{R,A}(\varepsilon, x), \hat G^{R,A} (\varepsilon, x)
\right]
=0, \quad
\left(\hat G^{R,A} (\varepsilon, x) \right)^2 =1.
\label{usadel-therm}
\end{equation}
Eq. \eqref{usadel-therm} can easily be solved with the result
\begin{equation}
\hat G^{R,A} = \pm \dfrac{\hat B^{R,A}}{\sqrt{(\hat B^{R,A})^2}},
\end{equation}
where $\hat B^{R,A}$ define a traceless part of the matrix $\hat \Omega^{R,A}$,
\begin{gather}
\hat B^{R,A} = \hat \Omega^{R,A} - \hat 1 \Omega_0^{R,A}, \quad \Sp(\hat B^{R,A})=0,
\\
\Omega_0^{R,A} =\dfrac{1}{4}\Sp(\hat \Omega^{R,A}).
\end{gather}
For simplicity here we assume that our system is spin isotropic and, hence, the matrix $\hat \Omega^{R,A}$ has no structure in the spin space.
Provided the superconducting order parameter $\Delta$ is purely real, the matrix Green functions have the following structure
\begin{equation}
\hat G^{R,A}=
\begin{pmatrix}
G^{R,A} & F^{R,A}
\\
-F^{R,A} & -G^{R,A}
\end{pmatrix},
\label{greenra-thermo}
\end{equation}
and the combination $\Sp( \hat \tau_3 \hat g_x^K)$ in Eq. (\ref{current-us1}) can be evaluated in a straightforward manner. As we already pointed out, the terms with higher derivatives in Eq. \eqref{gx-1} can be safely omitted provided the temperature gradient remains sufficiently small. With this in mind and making use of Eqs. \eqref{gk-thermo}-\eqref{greenra-thermo} we obtain
\begin{equation}
\Sp( \hat \tau_3 \hat g_x^K)
=
i\dfrac{\ell^2}{v_F}
\left(\Omega_0^R - \Omega_0^A\right)
\dfrac{4\varepsilon\nu(\varepsilon)\partial_x
T}{T^2\cosh^2\left[\varepsilon/(2T)\right]},
\end{equation}
where $\nu(\varepsilon) = (1/2)[G^R - G^A]$  is the density of states.
Substituting this combination into Eq. \eqref{current-us1}, we immediately arrive at the
expression for the thermoelectric coefficient
\begin{equation}
\alpha=
\dfrac{\sigma_N \ell}{2 e v_F}
\int d \varepsilon
\dfrac{\varepsilon \nu(\varepsilon)\Img\Omega_0^R}{
T^2\cosh^2\left[\varepsilon/(2T)\right]}.
\label{alpha}
\end{equation}
Note that an attempt to evaluate the expression (\ref{alpha}) making use of the standard Usadel equations immediately yields an incorrect result $\alpha =0$. This is because in local equilibrium one has $\Sp (\hat \tau_3 \check G
\partial_x \check G)^K=0$ and, hence, the contribution to the electric current from the
first term in the right-hand side of Eq. \eqref{gx-1} vanishes identically. In order to recover the correct expression for the thermoelectric coefficient it is necessary to go beyond the Usadel approximation and to include gradient corrections evaluated in the previous section.
E.g., in the case of superconductors doped with randomly distributed magnetic impurities \cite{Rusinov69,Kalenkov12} these correction terms allow to properly account for the electron-hole imbalance effects. For $\Omega_0^R \neq
\Omega_0^A$ Eq. \eqref{alpha} yields a non-zero result which -- in the limit $\ell \ll \xi_0$ considered here -- exactly coincides with the expression for $\alpha$ derived within a somewhat different approach \cite{Kalenkov12}. The corresponding calculation
is already presented in Ref. \onlinecite{Kalenkov12}, therefore it is not necessary to go into further details here.

\section{Photoelectric effect}
\label{sec-photo}

Let us apply the formalism developed above to the description of the photoelectric effect in superconductors doped with magnetic impurities.
Here we will concentrate on a hybrid structure depicted in Fig. 1. This structure consists of a superconductor and a normal metal separated by
an insulating barrier with cross section $\mathcal{A}$ and normal state resistance $R_N$. We will assume that the superconducting part of this
SN structure is uniformly irradiated by an ac electromagnetic field and that the corresponding penetration depth for this field exceeds
the thickness of the S-film. In the presence of magnetic impurities a combined effect of an ac field and electron scattering at such impurities yields charge imbalance production in a superconductor \cite{Zaitsev82,Zaitsev86} which, in turn, implies the appearance of a nonvanishing dc voltage $\overline{V}$ across the insulating barrier. This effect can be detected, e.g., by measuring the dc current $\overline{I}=\overline{V}/R_N$ across the SN interface. The current across the SN interface with low transmission can be expressed in the form \cite{Zaitsev84}
\begin{equation}
I=
\dfrac{eN_0\mathcal{A}}{16}
\int d \varepsilon
\left<
|v_{x}|D(\bm{p}_F)
\Sp(\hat g^K - \hat g^R h_N + h_N \hat g^A)
\right>,
\label{current}
\end{equation}
where $\hat g^{R,A,K}$ are retarded, advanced and Keldysh Green-Eilenberger functions on a
superconducting part of the interface, $h_N(\varepsilon) = \tanh (\varepsilon /2T)$ is related
to the equilibrium (Fermi) distribution function for electrons and holes in the normal metal
and $D(\bm{p}_F)$ is the angle-dependent transmission of the tunnel barrier. In the diffusive limit the Green functions only weakly depend on the direction of the Fermi momentum and, hence, the current \eqref{current} can be rewritten in a much simpler form
\begin{equation}
I=\dfrac{1}{16eR_N}
\int d \varepsilon
\Sp (\hat g_0^K - \hat g_0^R h_N + h_N \hat g_0^A),
\label{current2}
\end{equation}
where $\hat g^{R,A,K}_0 \equiv \left<\hat g^{R,A,K}\right>$ indicate the momentum averaged
Green-Eilenberger functions and
\begin{equation}
\dfrac{1}{R_N}=e^2 N_0 \mathcal{A}
\left< |v_{x}| D(\bm{p}_F)\right>.
\end{equation}
Combining Eq. \eqref{current} with the expression for the electric potential
\begin{equation}
eV=
\dfrac{1}{16}
\int d \varepsilon
\left<\Sp \hat g^K \right>
=
\dfrac{1}{16} \int d \varepsilon \Sp \hat g_0^K
\label{voltage}
\end{equation}
and making use of the condition $\Sp \overline{\hat g_0^R}= \Sp \overline{\hat g_0^A}=0$ we immediately recover the Ohmic dependence $\overline{I}=\overline{V}/R_N$, where the dc photovoltage $\overline{V}$ should be determined from the solution of the quasiclassical equations. Here and below the overbar denotes time averaging over the period of the external electromagnetic field.
\begin{figure}
\centerline{\includegraphics[width=80mm]{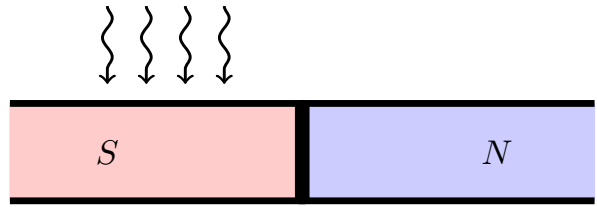}}
\caption{(Color online) Superconductor/normal metal tunnel junction irradiated
at microwave frequencies.}
\label{pe-sn-fig}
\end{figure}

The task at hand is to evaluate nonequilibrium Green functions $\hat g_0^{R,A,K}$ in a thin superconducting film in the presence of an ac electromagnetic field
which we describe by means of the time-dependent vector potential
\begin{equation}
\bm{A}(t)=\dfrac{\bm{E}c}{\omega}\cos\omega t,
\label{acem}
\end{equation}
where $\omega$ is the radiation frequency and $E$ denotes the amplitude
of the corresponding electric field component. Provided the intensity of the electromagnetic field is sufficiently small, i.e.
\begin{equation}
D\dfrac{\bm{E}^2 e^2}{\omega^2} \ll \max (T,\Delta ),
\label{intensity}
\end{equation}
this field can be treated as a small perturbation. Within this approximation it suffices to disregard the time dependence of the momentum averaged Green functions. Replacing $\nabla_i \Rightarrow i(e/c)A_i \hat \tau_3$ in Eq. \eqref{UsadelCorr} and averaging this equation over time, we obtain
\begin{multline}
\biggl[\check \Omega(\varepsilon)
+
iD\dfrac{e^2\bm{E}^2}{4\omega^2}
\hat \tau_3
\left\{
\check{G}(\varepsilon + \omega) + \check{G}(\varepsilon - \omega)
\right\}
\hat \tau_3
+
\dfrac{e^2\ell^2\bm{E}^2}{12\omega^2}
\\\times
\left\{
2 \check \Omega(\varepsilon)
-
\hat \tau_3 \check \Omega (\varepsilon + \omega) \hat \tau_3
-
\hat \tau_3 \check \Omega(\varepsilon - \omega) \hat \tau_3
\right\}
, \check G(\varepsilon)\biggr]=0.
\label{usadel-rf}
\end{multline}
Here we omitted terms proportional to $\bm{E}^4$ and for the sake of simplicity denoted time averages of the matrices $\check G$ and $\check \Omega$ respectively by the same symbols. According to Eq. \eqref{Omega}, the matrix $\check \Omega$ is expressed in terms of the energy $\varepsilon$, the order parameter $\Delta$ and the self-energy $\check \Sigma_{\text{m}}$. For a simple model of randomly distributed isotropic magnetic impurities the latter quantity reads \cite{Shiba68,Rusinov69}
\begin{multline}
\check{\Sigma}_{\text{m}}=\dfrac{n_{\text{imp}}}{2\pi N_0}
\Bigl\{
\left( [u_1 + \hat \tau_3 u_2]^{-1} +
i \check{g}_0 \right)^{-1}
\\+
\left( [u_1 - \hat \tau_3 u_2]^{-1} + i \check{g}_0 \right)^{-1}
\Bigr\},
\label{sigmam}
\end{multline}
where $n_{\text{imp}}$ is the concentration of magnetic impurities. Dimensionless parameters $u_1$ and $u_2$   characterize the strength of the spin-isotropic and spin-dependent part of the
impurity scattering potential. Following our convention we now employ the symbol $\check{g}_0$ to denote both time- and momentum-averaged Green-Eilenberger function matrix.

The relation between the functions $\check{g}_0$ and $\check G$ can be derived simply by averaging Eq. \eqref{G11} over time. As a result we obtain
\begin{multline}
\check{g}_0(\varepsilon)=
\check G(\varepsilon)
+\dfrac{\ell^2}{6}
\dfrac{e^2\bm{E}^2}{2\omega^2}
\biggl[
2\check{G}(\varepsilon)
\\-
\hat \tau_3 \check{G}(\varepsilon+\omega) \hat \tau_3
-
\hat \tau_3 \check{G}(\varepsilon-\omega) \hat \tau_3
\biggr],
\label{g-rf}
\end{multline}
Eqs. \eqref{usadel-rf}-\eqref{g-rf} fully determine a perturbative response of a superconductor to an external ac electromagnetic field (\ref{acem}).

Employing the normalization condition one can parameterize the Keldysh component of the
Green function matrix in a standard way as
\begin{equation}
\hat{G}^K = \hat{G}^R \hat h - \hat h \hat{G}^A,
\quad
\hat h = h_L + \hat \tau_3 h_T,
\end{equation}
where the functions $h_L$ and $h_T$ are directly linked to the distribution functions for electrons and holes. In equilibrium we have $h_L(\varepsilon) = \tanh (\varepsilon /2T)$ while the function $h_T$ equals to zero. In the presence of an ac field (\ref{acem}) the function $h_T$ already differs from zero and it can be found perturbatively from the kinetic equation derived by multiplying the Keldysh component of the Usadel equation \eqref{usadel-rf} by $\hat \tau_3$ and by taking the trace of the resulting expressions. This equation reads
\begin{multline}
\Img \left\{ \Delta^* [F^R(\varepsilon) + F^A(\varepsilon) ] \right\}
h_T(\varepsilon)
=
-\dfrac{\ell^2}{6}
\dfrac{e^2 \bm{E}^2}{2\omega^2}\nu(\varepsilon)
\\\times
\Biggl\{
\nu(\varepsilon+\omega)
\left[d(\varepsilon) + d(\varepsilon+\omega) \right]
\left[ h_L (\varepsilon) - h_L (\varepsilon + \omega)\right]
\\+
\nu(\varepsilon-\omega)
\left[d(\varepsilon) + d(\varepsilon-\omega) \right]
\left[ h_L (\varepsilon) - h_L (\varepsilon - \omega)\right]
\Biggr\},
\label{kin-rf}
\end{multline}
where $d(\varepsilon)$ is proportional to the diagonal part of the impurity self-energy
\begin{equation}
\Sp\left[\hat{\Sigma}_{\text{m}}^R(\varepsilon) - \hat{\Sigma}_{\text{m}}^A(\varepsilon)\right]
=
-4i\nu(\varepsilon)d(\varepsilon).
\end{equation}
The function $d(\varepsilon)$ depends on the impurity
scattering potential parameters $u_{1,2}$ as well as on the
quasiparticle energy $\varepsilon$. We find
\begin{equation}
d(\varepsilon)=
\dfrac{n_{\text{imp}}}{\pi N_0}
\dfrac{16 u_1 u_2^2 (1+u_1^2 - u_2^2) \Img G^R(\varepsilon)}{
\left| (1+u_1^2 - u_2^2)^2 + 4u_2^2 [G^R(\varepsilon)]^2 \right|^2}.
\end{equation}

Within our model electron scattering on magnetic impurities is responsible for
charge imbalance generation \cite{Zaitsev82,Zaitsev86} as well as for its relaxation \cite{Schmidt75}: These two effects are accounted for respectively by the left-hand side and the right-hand side of Eq. \eqref{kin-rf}. In order resolve this equation it suffices to set both the retarded and advanced Green functions equal to their bulk equilibrium values. The resulting photovoltage value $\overline{V}$ takes the form
\begin{equation}
e\overline{V}=\dfrac{1}{2}
\int d \varepsilon
\nu (\varepsilon) h_T (\varepsilon),
\label{photovoltage}
\end{equation}

\begin{figure}
\centerline{\includegraphics[width=80mm]{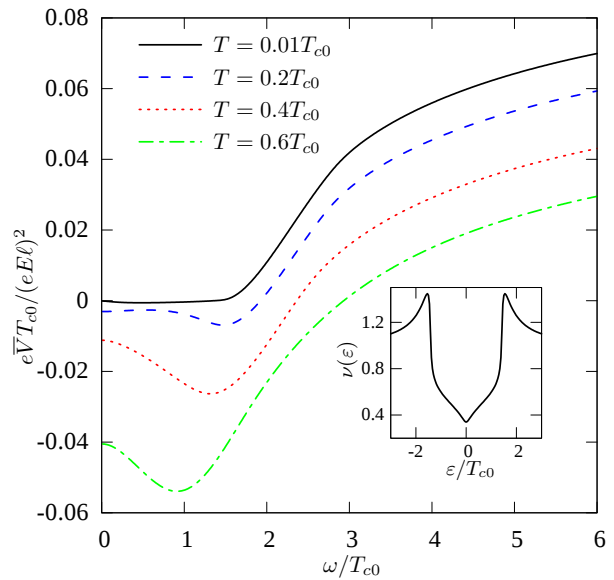}}
\caption{(Color online) Photovoltage $\overline{V}$ as a function of the microwave radiation frequency $\omega$. The magnetic impurity potential parameters ($u_1=u_2=1.0$) and impurity concentration ($n_{\text{imp}} = \pi N_0 T_{c0}$) are the same for all curves. The critical temperature of a superconductor in the absence of the magnetic impurities is denoted as $T_{c0}$. The inset shows zero temperature density of states as a function of energy. }
\label{pe-rf3-fig}
\end{figure}

In Fig. \ref{pe-rf3-fig} we illustrate typical frequency dependencies of the photovoltage  evaluated with the aid of Eqs. \eqref{kin-rf}-\eqref{photovoltage} at different temperatures. The magnetic impurity concentration and the impurity potential parameters are chosen to provide the regime of gapless superconductivity. In this regime the impurity bands overlap with the continuous spectrum  and the quasiparticle density of states remains nonzero in the whole energy range (see the inset in Fig. \ref{pe-rf3-fig}).
Quasiparticle transitions between the states in the continuous spectrum and/or the states in the impurity band may yield electron-hole imbalance of a different sign depending on the quasiparticle energy $\varepsilon$. The photovoltage value $\overline{V}$ is then controlled by an interplay between these processes and it may change the sign depending on the electromagnetic field frequency.

In general the voltage $\overline{V}$ is a complicated function of several parameters, such as temperature $T$, the magnetic impurity concentration $n_{\text{imp}}$ and the potential parameters $u_1$ and $u_2$ as well as the frequency of the external electromagnetic field $\omega$. Similarly to \cite{Kalenkov12} one can expect that the maximum value of $\overline{V}$ is achieved provided both temperature and the magnetic impurity concentration are approximately equal to one half of their respective critical values. In this case and at frequency values several times bigger than $T_{c0}$ (cf., e.g., Fig. 2) one can roughly estimate
\begin{equation}
e\overline{V} \sim \dfrac{\ell^2 e^2 \bm{E}^2}{T_{c0}}.
\label{phV}
\end{equation}
At the border of applicability of Eq. \eqref{intensity} one can set $\bm{E}^2 \sim T_{c0}^3/(e^2 v_F \ell)$ and obtain
\begin{equation}
e\overline{V} \sim T_{c0}^2 \tau_e,
\end{equation}
where $\tau_e=\ell /v_F$ is the elastic electron scattering time.

\section{Conclusions}
In this paper we extended the well known quasiclassical formalism of Usadel equations by deriving the leading gradient corrections to these equations.
At the first sight, such corrections should remain small in the diffusive limit and, hence, may at most yield subleading in the parameter $\ell /\xi_0$ results. Our analysis demonstrates
that it is not always so. If, for instance, magnetic impurities are present in a diffusive superconductor, spin-sensitive electron scattering at such impurities
introduces electron-hole asymmetry which results in anisotropic in the momentum corrections to the quasiparticle distribution function. Such corrections obviously cannot be
captured within the standard Usadel approximation but they are easily recovered with the aid of the formalism developed here.

It turns out that the presence of such anisotropic in the momentum terms in the distribution function may result in dramatic changes in the
behavior of a superconductor. For instance, the thermoelectric effect in superconductors (which is usually small in the parameter $T_c/\varepsilon_F$)
gets greatly enhanced in the presence of magnetic impurities and the related electron-hole imbalance in the system \cite{Kalenkov12}. We demonstrated
that this effect can be conveniently described employing the extension of the quasiclassical Usadel equation formalism worked out here.

Our formalism is also well suited in order to describe another interesting phenomenon -- the so-called photoelectric effect in superconductors and SN hybrid structures which implies the appearance of a dc voltage drop in the system as a result of inhomogeneous absorption of light. Also this effect is strongly enhanced provided a certain concentration of magnetic impurities is present in a superconductor. If such a superconductor in an SN system is irradiated by an external ac electromagnetic field the charge imbalance develops which causes a non-vanishing dc voltage $\overline{V}$ to occur across a tunnel barrier at the SN interface. Here we evaluated the magnitude of this voltage and demonstrated that in the presence of magnetic impurities in a superconductor the maximum value of $\overline{V}$ is controlled by the electromagnetic signal and the electron elastic mean free path $\ell$ (cf. Eq. (\ref{phV})) rather than by the small parameter $T_c/\varepsilon_F$ which just drops out in this case. This large photoelectric effect in SN hybrid structures with magnetic impurities can easily be detected in modern experiments and can also be directly employed in various applications, such as, e.g., a new generation of ultrasensitive bolometers.
\vspace{10mm}

\section*{Acknowledgements}
This work was supported in part by RFBR Grant No. 15-02-08273.

\appendix
\section{}
\label{appendix}
Making use of Eqs. \eqref{eilen_i}, \eqref{norm_i}, \eqref{gik} and \eqref{G11} after a straightforward but rather lengthy calculation we arrive at the following expression for the matrix $\check g_i$ which includes the leading order correction to Eq. \eqref{g_iUsadel}:
\begin{widetext}
\begin{multline}
\check g_i
=
-\ell (\check G \nabla_i \check G - \nabla_i)
-i
\dfrac{\ell^2}{v_F}
\left(
\check \Omega \nabla_i \check G
-
\check \Omega \check G \nabla_i
+
\check G \nabla_i \check \Omega
-
\nabla_i \check G \check \Omega
\right)
-
\dfrac{\ell^3}{60}
\Biggl\{
30 \check G \nabla_k \check G \nabla_i \check G \nabla_k \check G
-
15 \check G \nabla_k \check G \nabla_k \check G \nabla_i \check G
-\\-
15 \check G \nabla_i \check G \nabla_k \check G \nabla_k \check G
-
11 \nabla_i \check G \nabla_k \nabla_k \check G
-
11 \check G \nabla_k \nabla_k \check G \nabla_i
+
29 \nabla_k \check G \nabla_k \nabla_i \check G
+
29 \check G \nabla_i \nabla_k \check G \nabla_k
+
9 \check G \nabla_k \check G \nabla_k \nabla_i
+
9 \nabla_i \nabla_k \check G \nabla_k \check G
-\\-
11 \nabla_i \check G \nabla_k \check G \nabla_k
-
11 \nabla_k \check G \nabla_k \check G \nabla_i
-
6 \nabla_k \nabla_i \check G \nabla_k \check G
-
6 \check G \nabla_k \check G \nabla_i \nabla_k
-
6 \nabla_k \check G \nabla_i \nabla_k \check G
-
6 \check G \nabla_k \nabla_i \check G \nabla_k
-
6 \nabla_k \check G \nabla_i \check G \nabla_k
+\\+
6 \nabla_k \nabla_i \nabla_k
-
6 \check G \nabla_k \nabla_i \nabla_k \check G
-
\check G \nabla_k \nabla_k \nabla_i \check G
-
\check G \nabla_i \nabla_k \nabla_k \check G
-
\nabla_k \nabla_k \check G \nabla_i \check G
-
\check G \nabla_i \check G \nabla_k \nabla_k
+
\nabla_i \nabla_k \nabla_k
+
\nabla_k \nabla_k \nabla_i
\Biggr\}
\label{giCorr}
\end{multline}
\end{widetext}
Employing this result together with Eq. \eqref{eilen_0} we obtain the resulting equation for the matrix function $\check G$:
\begin{widetext}
\begin{multline}
\left[ \check \Omega - i \dfrac{v_F\ell}{3}\nabla_i \check G \nabla_i, \check G \right]
-
\dfrac{\ell^2}{6}
\left[
\check \Omega \nabla_i \nabla_i
+
\nabla_i \nabla_i \check \Omega
-
2 \nabla_i \check \Omega \nabla_i, \check G \right]
+
i\dfrac{v_F\ell^3}{180}
\biggl[
-30 \nabla_i \check G \nabla_k \check G \nabla_i \check G \nabla_k
-
29 \nabla_i \nabla_k \check G \nabla_k \nabla_i
+\\+
15 \nabla_i \check G \nabla_k \check G \nabla_k \check G \nabla_i
+
15 \nabla_i \check G \nabla_i \check G \nabla_k \check G \nabla_k
+
11 \nabla_i \nabla_i \check G \nabla_k \nabla_k
+
6 \nabla_k \nabla_i \check G \nabla_k \nabla_i
+
6 \nabla_k \check G \nabla_i \nabla_k \nabla_i
+
6 \nabla_k \nabla_i \nabla_k \check G \nabla_i
+\\+
\nabla_i \nabla_k \nabla_k \check G \nabla_i
+
\nabla_i \check G \nabla_k \nabla_k \nabla_i
+
\nabla_i \check G \nabla_i \nabla_k \nabla_k
+
\nabla_i \nabla_i \nabla_k \check G \nabla_k
, \check G \biggr]
=0.
\label{UsadelCorr}
\end{multline}
\end{widetext}
The first commutator in Eq. \eqref{UsadelCorr} represents the standard Usadel equation and last two terms define small corrections.

It is easy to observe that within the approximations adopted here only the term $\check g_i n_i$ in the expansion \eqref{expansion}
contributes to the electric current \eqref{current-eil}. Then for the $i$-th component of the current density we have
\begin{equation}
j_i=
-\dfrac{eN_0v_F}{24}
\int d \varepsilon
\Sp\left(\hat g_i^K \hat \tau_3 \right),
\label{current-us}
\end{equation}
where $\hat g_i^K$ is the Keldysh component of the matrix \eqref{giCorr}.

All the equations presented here remain applicable as long as the matrix function $\check G$ changes at distances exceeding the mean free path $\ell$.
Provided $\check G$ changes in space sufficiently slowly one can neglect all terms in curly brackets in Eq. \eqref{giCorr} as well as the last commutator
in Eq. \eqref{UsadelCorr}. Then the above equations receive significant simplifications. Further simplifications of our formalism occur
in the absence of the magnetic field and provided the function $\check G$ depends only one one coordinate, see the corresponding discussion in Sec. \ref{subsec-corr}.

\end{document}